\documentclass[aps,prb,reprint,titlepage]{revtex4-2}

\usepackage{amsmath}  
\usepackage{amsfonts} 
\usepackage{graphicx} 
\usepackage[utf8]{inputenc}
\usepackage[T1]{fontenc}

\begin{document}

\title{Visualization-Based Approach to Condensed-Phase Line Broadening Using Polyene Chains}

\author{Saba Mahmoodpour}
\affiliation{Department of Chemistry, University of North Carolina at Chapel Hill, Chapel Hill, NC 27599, USA}

\author{Andrew M. Moran}
\email{ammoran@unc.edu}
\affiliation{Department of Chemistry, University of North Carolina at Chapel Hill, Chapel Hill, NC 27599, USA}

\begin{abstract}
Condensed-phase spectral line shapes encode the strength and timescale of interactions between molecules and their environments, yet these ideas are often difficult to introduce at the undergraduate level due to their reliance on formal theoretical treatments. We present a visualization-based approach that combines analytic results with numerical simulations to illustrate the physical origins of spectral line broadening in conjugated molecular systems. Using a time-dependent H\"uckel Hamiltonian, we derive closed-form expressions for coherent electronic motion in finite polyene chains and show how these results provide direct insight into the role of molecular orbital structure in light absorption. Environmental effects are introduced through stochastic fluctuations of the Hamiltonian matrix elements, allowing students to observe how system--environment interactions disrupt coherent motion and produce scattering-like features in electronic trajectories. Real-space animations and simulated absorption spectra provide an intuitive link between microscopic dynamics and measured line shapes. The MATLAB code provided with this work offers an accessible platform for integrating computation and visualization into undergraduate instruction while introducing key concepts in condensed-phase spectroscopy.
\end{abstract}


\maketitle 

\section{Introduction}
Undergraduate physical chemistry courses motivate the study of systems and phenomena at the interface of chemistry and physics \cite{Atkins2018V2,EngelReid2018}. This subdiscipline has a long and successful tradition of introducing spectroscopy through gas-phase rotational and vibrational spectra, where sharp spectral features allow students to connect quantum energy levels directly to experimental observables \cite{williamsonTeachingRovibronicSpectroscopy2007,bayramRotationalSpectraN22015,kashemMolecularSpectroscopyLaboratory2023}. These systems are well-suited for developing intuition about selection rules, transition frequencies, and lifetime broadening arising from spontaneous emission. However, much of contemporary chemical research focuses on condensed-phase systems, such as liquids and solids, where spectral lines are rarely sharp and instead carry important dynamical information in their widths and shapes. In these environments, line broadening reflects interactions with a surrounding medium and encodes the microscopic fluctuations that underlie functional processes such as energy transport, electron transfer, and charge conduction \cite{mukamelPrinciplesNonlinearOptical1995,nitzanChemicalDynamicsCondensed2014,Cina2022}. Although the teaching of condensed-phase spectroscopy has traditionally been challenged by heavy formalism, advances in computational power and visualization tools now make it possible to introduce these concepts at the undergraduate level in a concrete and intuitive way.

Here, we present a computational approach in which visualization plays a central role in making these abstract ideas tangible for students. Rather than developing the full model analytically, we combine closed-form results for coherent electronic motion with simulations that model the interaction of light with a polyene chain undergoing stochastic fluctuations in real space. Time-dependent disorder leads to scattering-like events that appear as interruptions in the motion of an electronic wavepacket, providing a direct visual manifestation of decoherence. Students can compare individual trajectories, which reveal how phase coherence is gradually lost through repeated interactions with an effective environment, with ensemble-averaged inhomogeneous line broadening in the frequency domain. Moreover, tuning a single parameter that controls the strength of system--environment coupling enables a continuous transition from coherent, reversible motion to strongly dephased dynamics and broadened absorption spectra. This visual and interactive framework allows students to connect time-domain dynamics with frequency-domain line shapes, thereby reinforcing core concepts while lowering the barrier to engaging with modern condensed-phase spectroscopy.

The theoretical foundations required to understand line broadening are well established, but they must be carefully navigated in an undergraduate course that is not specialized in spectroscopy or open quantum systems. Introductory treatments typically begin with the classical forced-oscillator model, which provides an intuitive picture of light absorption, resonance, and damping. Translating this picture to a quantum molecular system involves recognizing that absorption corresponds to the coherent motion of electronic charge, a viewpoint that can be captured using simple molecular orbital models. The H\"uckel tight-binding model is particularly well suited for this purpose because it is familiar to chemistry students and captures essential features of electronic structure and delocalization \cite{Huckel1931,RobertsHuckelModel1960,StreitwieserHuckelBook1961}. In condensed-phase environments, molecular Hamiltonians are not static but fluctuate due to interactions with their surroundings. Rather than explicitly modeling the environment, these effects are incorporated here through a reduced description in which disorder and time-dependent fluctuations are introduced directly into the molecular Hamiltonian. This approach preserves the essential physics of environment-induced dephasing while remaining  accessible to undergraduates.

This work aims to provide a physically intuitive framework for understanding how electronic coherence, environmental fluctuations, and optical line broadening are connected in conjugated molecular systems. The presentation emphasizes simple models, analytic solutions for coherent dynamics, explicit time-domain dynamics, and direct connections to familiar oscillator-based descriptions of the optical response. This perspective is intended to help students and instructors build intuition for condensed-phase spectroscopy while retaining a rigorous microscopic foundation. To support practical classroom adoption, the accompanying MATLAB code is designed with a user-friendly graphical interface, enabling students and instructors to explore the influence of microscopic fluctuations on the optical response across a range of model parameters. In-class polling problems employing this approach have been developed and used in upper-level undergraduate and graduate physical chemistry courses. Example problem set are provided in the Supplementary Material to facilitate direct implementation of these materials in instructional settings.

\section{Modeling Line Broadening Through Dynamic Disorder}

Spectral line broadening in condensed-phase systems arises from stochastic energy level fluctuations induced by interactions between chromophores and their local environments. In the present model, these effects are incorporated through a time-dependent electronic Hamiltonian, allowing students to visualize how microscopic fluctuations disrupt coherent electronic motion and lead to broadened absorption spectra. Importantly, two distinct physical mechanisms are captured: (i) dynamic disorder acting within individual trajectories, which produces decoherence and scattering-like dynamics in the time domain, and (ii) ensemble averaging over many realizations, which gives rise to inhomogeneous broadening in the frequency domain.

Under resonant excitation, the optical response of a conjugated molecule is dominated by a single characteristic frequency associated with the lowest-energy electronic transition. As shown below, the light absorption process reflects coherent motion of the electronic probability distribution along the molecular backbone. This observation provides a direct connection between the present numerical propagation approach and the classical driven Lorentz oscillator model that is widely used to describe optical absorption and dispersion in molecular systems. In this familiar picture, light drives a bound electronic coordinate that responds harmonically at the resonant frequency, while environmental interactions introduce damping and energy dissipation. The time-dependent Hamiltonian framework adopted here may therefore be viewed as a microscopic realization of this oscillator-based description, in which the electronic motion is treated quantum mechanically and environmental effects are incorporated through stochastic fluctuations of the underlying electronic parameters.

\subsection{The H\"uckel Model for Electronic Dynamics}

In this work, the H\"uckel model serves as the starting point for simulating light-driven electronic dynamics in a molecular chain. This approach provides a simple and physically transparent framework for describing the electronic structure of conjugated molecular systems and is well-suited for pedagogical applications. In its standard form, the model represents the $\pi$-electron system of a linear polyene as a set of coupled atomic sites, each corresponding to a carbon $p_z$ orbital. The electronic Hamiltonian is written as a matrix $\mathbf{H}$, whose elements encode both the local electronic energies and the electronic couplings between neighboring sites. 

In matrix form, the Hamiltonian is written as
\begin{equation}
\mathbf{H} =
\begin{pmatrix}
\alpha_1 & \beta_{12} & 0 & \cdots & 0 \\
\beta_{21} & \alpha_2 & \beta_{23} & \cdots & 0 \\
0 & \beta_{32} & \alpha_3 & \cdots & 0 \\
\vdots & \vdots & \vdots & \ddots & \beta_{N-1,N} \\
0 & 0 & 0 & \beta_{N,N-1} & \alpha_N
\end{pmatrix},
\end{equation}
where $\alpha_k$ represents the energy of the $k$th carbon atom and $\beta_{km}$ is the electronic coupling between neighboring sites $k$ and $m$. Physically, the diagonal elements $\alpha_k$ set the effective energy level for occupying the $p_z$ orbital on atom $k$, whereas the off-diagonal elements $\beta_{km}$ quantify the electronic coupling between neighboring atoms along the conjugated backbone. In the absence of disorder, these parameters are typically taken to be uniform along the chain, reflecting the chemical equivalence of the repeating units.

The molecular orbitals are obtained by solving the matrix eigenvalue equation,
\begin{equation}
\mathbf{H}\mathbf{C} = \mathbf{C}\mathbf{E},
\end{equation}
where $\mathbf{C}$ contains the molecular orbital coefficients and $\mathbf{E}$ is a diagonal matrix of the corresponding energy eigenvalues. For a linear polyene, the resulting molecular orbitals span a range of energies, with the highest occupied molecular orbital (HOMO) and lowest unoccupied molecular orbital (LUMO) defining the dominant optical transition in the visible or near-visible region of the spectrum. 

To model condensed-phase environments, the Hamiltonian is allowed to fluctuate in time through the introduction of dynamic disorder. Rather than treating the surrounding medium explicitly, environmental effects are incorporated by allowing both diagonal and off-diagonal matrix elements to vary stochastically around their mean values of $\alpha=-6.8$~eV and $\beta=-3.6$~eV \cite{EngelReid2018}. These fluctuations represent the influence of molecular vibrations, solvent motions, and other environmental degrees of freedom that perturb the electronic structure. Importantly, this approach preserves the simplicity of the H\"uckel framework while providing a microscopic description of environment-induced dephasing and spectral line broadening.

\subsection{Calculating Electron Trajectories on Polyene Chains}

The electronic state of the $\pi$ system is described in a basis of $N$ conjugated carbon atoms by the vector
\begin{equation}
\phi(t) =
\begin{pmatrix}
\phi_1(t) \\
\phi_2(t) \\
\vdots \\
\phi_N(t)
\end{pmatrix},
\end{equation}
where $\phi_k(t)$ denotes the probability amplitude for finding the electron on atom $k$ at time $t$. 
The square modulus $|\phi_k(t)|^2$ represents the instantaneous probability of locating the electron at each atomic site and is visualized directly in our real-space simulations. The time evolution of the electronic state is governed by the time-dependent Schr\"odinger equation
\begin{equation}
i\hbar \frac{d}{dt}\phi(t) = \mathbf{H}(t)\,\phi(t),
\label{schrodinger}
\end{equation}
where $\mathbf{H}(t)$ is the H\"uckel Hamiltonian. 
To model condensed-phase environments, the matrix elements of $\mathbf{H}(t)$ are allowed to undergo stochastic fluctuations in time, thereby capturing the effects of solvent dynamics and torsional motions of the polyene chain on the local site energies and electronic couplings.

To propagate the state forward in time, we first diagonalize the Hamiltonian for a specific realization of the stochastically varying matrix elements at time $t$:
\begin{equation}
\mathbf{H}(t)\mathbf{C}(t) = \mathbf{C}(t)\mathbf{E}(t),
\end{equation}
where the columns of $\mathbf{C}(t)$ are the instantaneous molecular orbitals and $\mathbf{E}(t)$ contains their energies.
Because the initial condition corresponds to a localized, non-stationary state (e.g., occupation of an atomic orbital on one end of the chain), the probability amplitudes can be represented as a superposition of instantaneous molecular orbitals at time $t$, with the expansion coefficients collected in the vector $\psi(t)$. At each time step, the vector of site-basis amplitudes $\phi(t)$ obtained by solving Equation~\eqref{schrodinger} is transformed into the molecular-orbital basis through the time-local matrix $\mathbf{C}(t)$:
\begin{equation}
\psi(t) = \mathbf{C}^{-1}(t)\,\phi(t).
\end{equation}

Under the assumption that the incident light source is resonant with the HOMO–LUMO absorption spectrum, the electronic dynamics are dominated by the highest occupied and lowest unoccupied molecular orbitals, resulting in periodic motion at a frequency set by the HOMO–LUMO energy gap. For a finite linear chain, the molecular orbitals are nondegenerate, and optical transitions involving other orbital pairs are spectrally detuned from the HOMO–LUMO resonance. In the weak-excitation regime relevant to light absorption, these off-resonant channels contribute negligibly to the dynamics. Formally, the time-evolution operator associated with the Schr\"odinger equation may be restricted to the subspace spanned by the HOMO and LUMO, while retaining the full spatial representation of the electronic wavefunction.

This restriction is implemented by selecting from $\mathbf{C}(t)$ the two columns corresponding to the HOMO and LUMO, forming the reduced matrix
\begin{equation}
\mathbf{C}_{\mathrm{HL}}(t)=\mathbf{C}(t)\,\mathbf{S}_{\mathrm{HL}},
\end{equation}
where $\mathbf{S}_{\mathrm{HL}}=(\mathbf{e}_{\mathrm{H}}\ \mathbf{e}_{\mathrm{L}})$ is a column-selection matrix that extracts the HOMO and LUMO columns of $\mathbf{C}(t)$, and $\mathbf{e}_\mathrm{H}$ and $\mathbf{e}_\mathrm{L}$ are standard unit vectors.
The associated diagonal energy matrix is given by
\begin{equation}
\mathbf{E}_{\mathrm{HL}}(t) =
\begin{pmatrix}
E_{\mathrm{H}}(t) & 0 \\
0 & E_{\mathrm{L}}(t)
\end{pmatrix}.
\end{equation}
Within this reduced molecular-orbital basis, the short-time propagator retains a diagonal form,
\begin{equation}
\psi_{\mathrm{HL}}(t+\Delta t) =
\exp\!\left[-\frac{i}{\hbar}\mathbf{E}_{\mathrm{HL}}(t)\Delta t\right]\psi_{\mathrm{HL}}(t),
\end{equation}
where $\psi_{\mathrm{HL}}(t)$ contains the probability amplitudes associated with the HOMO and LUMO only. Transforming back to the site basis yields a single effective propagation step,
\begin{equation}
\phi(t+\Delta t) =
\mathbf{C}_{\mathrm{HL}}(t)
\exp\!\left[-\frac{i}{\hbar}\mathbf{E}_{\mathrm{HL}}(t)\Delta t\right]
\mathbf{C}_{\mathrm{HL}}^{\mathrm T}(t)\,\phi(t).
\end{equation}
This expression defines a site-basis propagator that evolves the electronic state entirely within the HOMO--LUMO subspace while preserving the full spatial extent of the wavefunction. Accordingly, the $N$ elements of the vector $\phi(t)$ represent the probability amplitudes for all carbon atoms in the polyene chain.

To incorporate environmental effects within this propagation scheme, the Hamiltonian is allowed to fluctuate in time through stochastic variations of its matrix elements,
\begin{equation}
\alpha_k(t) = \bar{\alpha}  + \delta\alpha_k(t),
\qquad
\beta_{km}(t) = \bar{\beta}  + \delta\beta_{km}(t),
\label{eq:dynamic_disorder_preview}
\end{equation}
where $\bar{\alpha}$ and $\bar{\beta}$ denote mean values, and $\delta\alpha_k(t)$ and $\delta\beta_{km}(t)$ represent time-dependent perturbations arising from nuclear motion, solvent fluctuations, and conformational dynamics of the polyene backbone.
At each time step, the electronic state is propagated using the instantaneous Hamiltonian \(\mathbf{H}(t)\), allowing environmental interactions to be incorporated into the time evolution.
The physical consequences of this dynamic disorder, including decoherence and the emergence of scattering-like behavior in individual trajectories, may be visualized using numerical methods.

\subsection{Harmonic Electronic Motion and Its Role in the Optical Response of Polyenes}
\label{analyticsection}

Before introducing the numerical simulations, it is useful to consider an idealized physical picture of how an electronic excitation moves along a conjugated molecular chain and how this motion connects to the optical response. Although the electronic dynamics of polyenes are inherently quantum mechanical, their real-space behavior under resonant optical excitation can be understood in surprisingly intuitive terms. In this section, we show that when a polyene is excited near its lowest-energy optical transition, the resulting electronic motion along the backbone is governed by a single characteristic frequency. This perspective provides a transparent link between microscopic electronic structure, time-dependent quantum dynamics, and the classical oscillator models that are widely used to introduce light absorption and dispersion. Establishing this connection clarifies the physical implications of the assumptions underlying the numerical propagation schemes employed below.

Within the Hückel model, the molecular orbitals of a linear polyene form
standing-wave patterns of the electron density.  For a chain of $N$
conjugated carbon atoms, the orbital coefficients and corresponding
$\pi$-electron energies are
\begin{equation}
C_{ku} = \sqrt{\frac{2}{N+1}}\,
\sin\!\left(\frac{u\pi k}{N+1}\right)
\label{eq:Cku}
\end{equation}
and
\begin{equation}
E_u = \alpha + 2\beta
\cos\!\left(\frac{u\pi}{N+1}\right),
\label{eq:Em}
\end{equation}
with $u = 1,2,\ldots,N$.  
Each molecular orbital therefore represents a stationary
standing wave with $u-1$ nodes distributed along the chain. When the electron is initially localized on one atomic site, its time evolution can be expressed as a coherent superposition of these stationary states. The resulting quantum interference produces the characteristic real-space motion of the electron along the polyene backbone.

When the incident light is resonant with the HOMO--LUMO transition,
the dynamics simplify dramatically.  Only the HOMO and LUMO
eigenstates, with indices $\mathrm{H}=N/2$ and $\mathrm{L}=N/2+1$ for a neutral polyene,
contribute significantly to the evolution.  The site-basis amplitude on
atom $k$ is then
\begin{equation}
\phi_k(t)
=
C_{1\mathrm{H}}\, C_{k\mathrm{H}}\, e^{-i E_{\mathrm{H}} t/\hbar}
+
C_{1\mathrm{L}}\, C_{k\mathrm{L}}\, e^{-i E_{\mathrm{L}} t/\hbar},
\label{eq:main_phi_HL}
\end{equation}
containing a single oscillation frequency determined by the
HOMO--LUMO energy gap.  For an electron initiated at the first site index, the probability of finding the electron on site $k$ at time $t$ becomes $P_{1\to k}(t) = |\phi_k(t)|^{2}$.
As shown in the Supplementary Material, the following compact expression can be obtained for the special case of the HOMO--LUMO pair of a neutral polyene with an even number of carbon atoms:
\begin{equation}
\begin{aligned}
P_{1\to k}(t)
&=
\frac{4}{(N+1)^2}
\cos^{2}\!\left(\frac{\pi}{2(N+1)}\right)
\\[4pt]
&\quad \times
\Bigg[
1
- \cos(k\pi)\cos\!\left(\frac{k\pi}{N+1}\right)
\\[4pt]
&\qquad
- \cos(k\pi)\cos\!\left(\frac{E_{\mathrm{L}} - E_{\mathrm{H}}}{\hbar}\, t\right)
\\[4pt]
&\qquad
+ \cos\!\left(\frac{k\pi}{N+1}\right)
  \cos\!\left(\frac{E_{\mathrm{L}} - E_{\mathrm{H}}}{\hbar}\, t\right)
\Bigg].
\end{aligned}
\label{eq:rabichain}
\end{equation}
This formula shows that the electron undergoes strictly periodic,
Rabi-like motion along the chain, since only a single frequency
$(E_{\mathrm{L}} - E_{\mathrm{H}})/\hbar$ governs the dynamics.  In contrast, propagation in the
full $N$-orbital space incorporates all molecular-orbital frequencies and
exhibits a complex revival structure.

Notably, for ethylene ($N=2$), the analytic expressions simplify to
\begin{equation}
C_{11} = C_{21} = \frac{1}{\sqrt{2}},
\qquad
C_{12} = -C_{22} = \frac{1}{\sqrt{2}},
\end{equation}
and the HOMO--LUMO energy separation becomes $(E_{\mathrm{L}} - E_{\mathrm{H}}) = 2\beta$.  The probability of finding the electron on the second carbon
atom is then
\begin{equation}
P_{1\to 2}(t)
=
\sin^{2}\!\left(\frac{\beta t}{\hbar}\right),
\label{eq:main_Rabi_ethylene}
\end{equation}
which is the classic Rabi solution for a two-level system.  The electron
oscillates between the two carbon atoms,
illustrating the simplest manifestation of HOMO--LUMO-driven electron
motion.

An instructive way to summarize this behavior is to examine the mean position of the electron along the chain.  
Defining the mean site index as \(\langle k\rangle(t)=\sum_k k\,P_{1\to k}(t)\), it can be shown that the probability distribution undergoes simple harmonic motion,
\begin{equation}
\langle k\rangle(t)=k_0+\Delta k\,\cos(\Omega t),
\end{equation}
where \(\Omega=(E_{\mathrm{L}} - E_{\mathrm{H}})/\hbar\) is the HOMO--LUMO beat frequency and \(\Delta k\) is the oscillation amplitude
(as derived in the Supplementary Material).  
The mean ``velocity'' corresponding to this coordinate is given by the time derivative,
\begin{equation}
v(t)=\frac{d}{dt}\langle k\rangle(t)=-\,\Omega\,\Delta k\,\sin(\Omega t),
\end{equation}
which is phase-shifted by \(\pi/2\) relative to the mean position, as in a classical, driven harmonic oscillator.
Thus, although the underlying dynamics arise from quantum interference between two stationary molecular orbitals,
the resulting real-space motion of the electron along the polyene backbone closely mirrors that of a particle oscillating in a quadratic potential.  

\begin{figure*}[!t]
\centering
\includegraphics[width=1.0\linewidth]{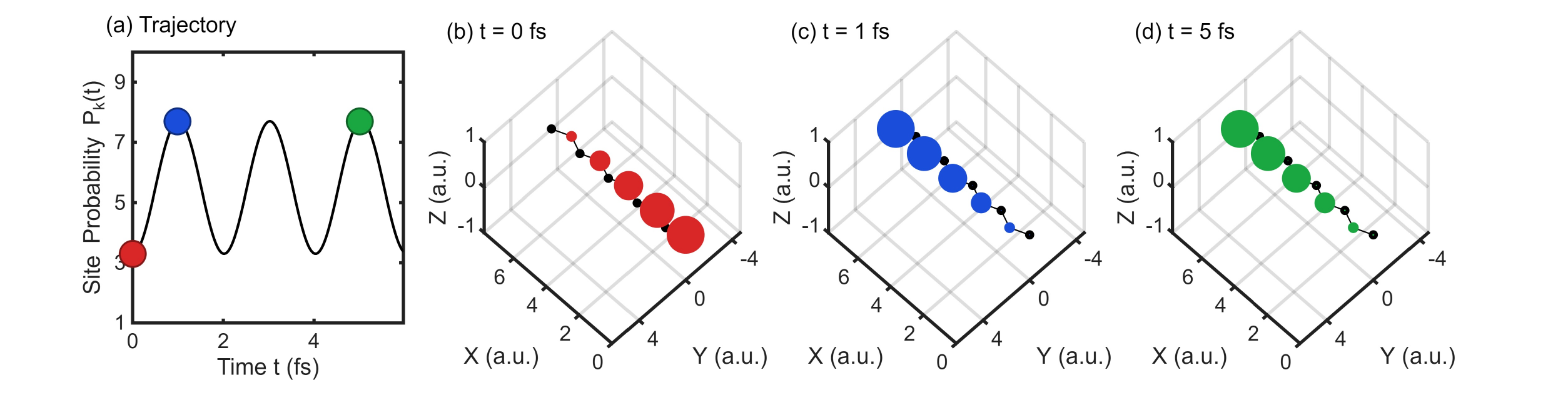}
\caption{Coherent electronic dynamics are computed for a 10-membered polyene chain with a tridiagonal H\"uckel Hamiltonian.
(a) Time-dependent expectation value of the carbon-site index computed with diagonal ($\delta\alpha$) and off-diagonal ($\delta\beta$) fluctuations set equal to zero. (b)--(d) Real-space snapshots of the electronic probability distribution evaluated at successive turning points of the trajectory at $t = 0$, 1, and 5~fs, respectively. In the absence of stochastic fluctuations, the electron dynamics are strictly periodic, leading to exact recurrences in the probability distribution.}
\label{closedsystem}
\end{figure*}

The emergence of this strictly harmonic motion also provides a clear physical justification for propagating the electronic dynamics within the resonant HOMO--LUMO subspace.
From this perspective, the HOMO--LUMO time evolution represents the microscopic quantum origin of the classical
Lorentz oscillator model that is traditionally used as the starting point for describing optical absorption
and dispersion in conjugated systems.  At the level of the mean electronic motion, the dynamics may be viewed
as obeying an effective driven-oscillator equation of the form
\begin{equation}
\frac{d^{2}}{dt^{2}}\langle k\rangle(t)
+\Omega^{2}\langle k\rangle(t)
\;\propto\; E(t),
\label{eq:effective_lorentz_k}
\end{equation}
where \(E(t)\) is the applied electric field and the proportionality reflects the coupling of the field to the
electronic polarization along the chain.  The restriction of the time-evolution operator to the HOMO--LUMO
manifold captures the essential resonant degree of freedom that dominates the optical response,
providing a natural and physically transparent foundation for later extensions that incorporate driving,
damping, and environmental effects.
    
\section{Calculated Polyene Optical Response and Absorption Line Shapes}

In this section, we first present numerical simulations of the electronic dynamics in an idealized polyene chain without environmental fluctuations. This limit provides a clear baseline where the motion is fully coherent and strictly periodic, allowing the connection between molecular orbital structure and real-space dynamics to be seen without additional complications. We then introduce stochastic fluctuations into the Hamiltonian to model interactions with a condensed-phase environment and examine how these perturbations modify the dynamics and the resulting absorption spectra. By comparing simulations with and without fluctuations, we show that electronic coherence is far more sensitive to disorder in the inter-site couplings than to disorder in the site energies. This contrast provides physical insight into the distinct microscopic origins of spectroscopic line broadening viewed from a time-domain perspective.

The calculations presented in Figure~\ref{closedsystem} represent the behavior of the idealized polyene chain described in Section~\ref{analyticsection}. The dynamics are initiated from an initial condition in which the electron fully populates an atomic orbital at one end of the polyene chain. The site probabilities then evolve according to Equation~\eqref{eq:rabichain}, revealing the sinusoidal nature of the electronic trajectory under resonant conditions.  The site probabilities are plotted at opposite turning points at $t=0$ and $1$~fs, consistent with the oscillation period of approximately 2~fs. As required by Equation~\eqref{eq:rabichain}, the probability distribution precisely recurs every cycle, yielding identical results at $t=1$ and $5$~fs. This coherent baseline provides the reference against which the effects of stochastic fluctuations may be assessed.

\begin{figure*}[ht]
\centering
\includegraphics[width=1.0\linewidth]{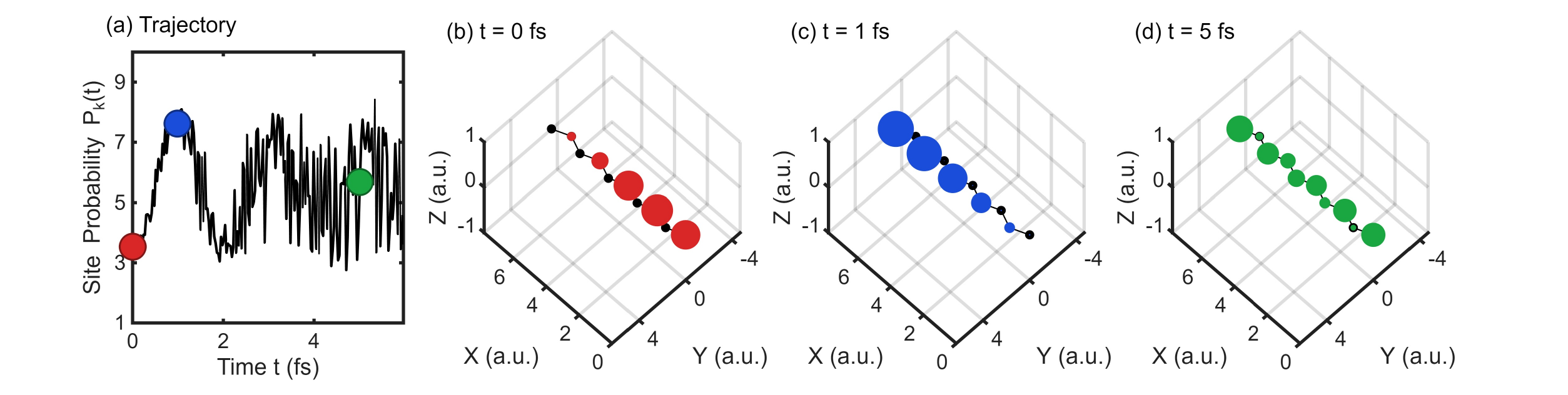}
\caption{Fluctuations induce scattering and decoherence in electronic trajectories along a polyene chain.
(a) Time-dependent expectation value of the carbon atom index computed with diagonal fluctuations set to $\delta\alpha = 0.00$~eV and off-diagonal fluctuations $\delta\beta = 0.06$~eV. (b)--(d) Real-space snapshots of the electronic probability distribution evaluated at successive turning points of the trajectory at $t = 0$, 1, and 5~fs, respectively. Repeated scattering events arising from dynamic fluctuations of the electronic couplings disrupt phase coherence, leading to a progressive loss of coherent electronic motion along the chain.}
\label{opensystem}
\end{figure*}

To illustrate the microscopic origin of line broadening, the site probability dynamics are computed for an individual trajectory with $\delta\alpha = 0.00$~eV and $\delta\beta = 0.06$~eV, as shown in Figure~\ref{opensystem}. Deviations between the idealized and stochastic trajectories increase as time evolves due to the accumulation of scattering events, which appear as weak reflections when the process is viewed as an animation. Progressive randomization of the electronic phase leads to decoherence over the first few cycles of the wavepacket, manifesting as spatial broadening of the wavepacket and loss of periodicity at $t = 1$ and $5$~fs. This behavior reflects the fact that the fluctuation amplitudes in the present model place the system in a strong-disorder regime, providing a natural transition to the discussion of fluctuation timescales, line-shape limits, and their physical interpretation. Viewed in this way, the animation clarifies that spectral line broadening emerges from the cumulative loss of the electron's phase coherence caused by many small, spatially localized scattering (reflection) events that are readily identifiable in the real-space dynamics, thereby connecting the familiar frequency-domain concept of linewidth directly to the underlying physics.

While the stochastic fluctuations are taken to be uncorrelated at successive time points in the present simulations, finite relaxation rates can be incorporated through an Ornstein--Uhlenbeck model for molecule--solvent interactions (a standard way of introducing finite ``memory'' into random fluctuations). Notably, molecular line-broadening dynamics in solutions at ambient temperature often conform to the slow-modulation regime of Kubo line-shape theory, where fluctuations evolve slowly compared to the electronic motion and lead to quasi-static energetic disorder \cite{mukamelPrinciplesNonlinearOptical1995}. In this limit, the spectral line shapes are Gaussian, while the absorption and fluorescence spectra exhibit a Stokes shift equal to twice the Marcus reorganization energy. Although the present implementation employs delta-correlated noise for numerical simplicity, the relatively large fluctuation amplitudes place the model in this strong-disorder, quasi-static limit rather than in the motional-narrowing regime (where very rapid fluctuations average out their effects). The resulting dynamics capture the key qualitative features of slow-modulation behavior, including strong dephasing and Gaussian-like broadening.

\begin{figure}[ht]
    \centering
\includegraphics[width=0.95\linewidth]{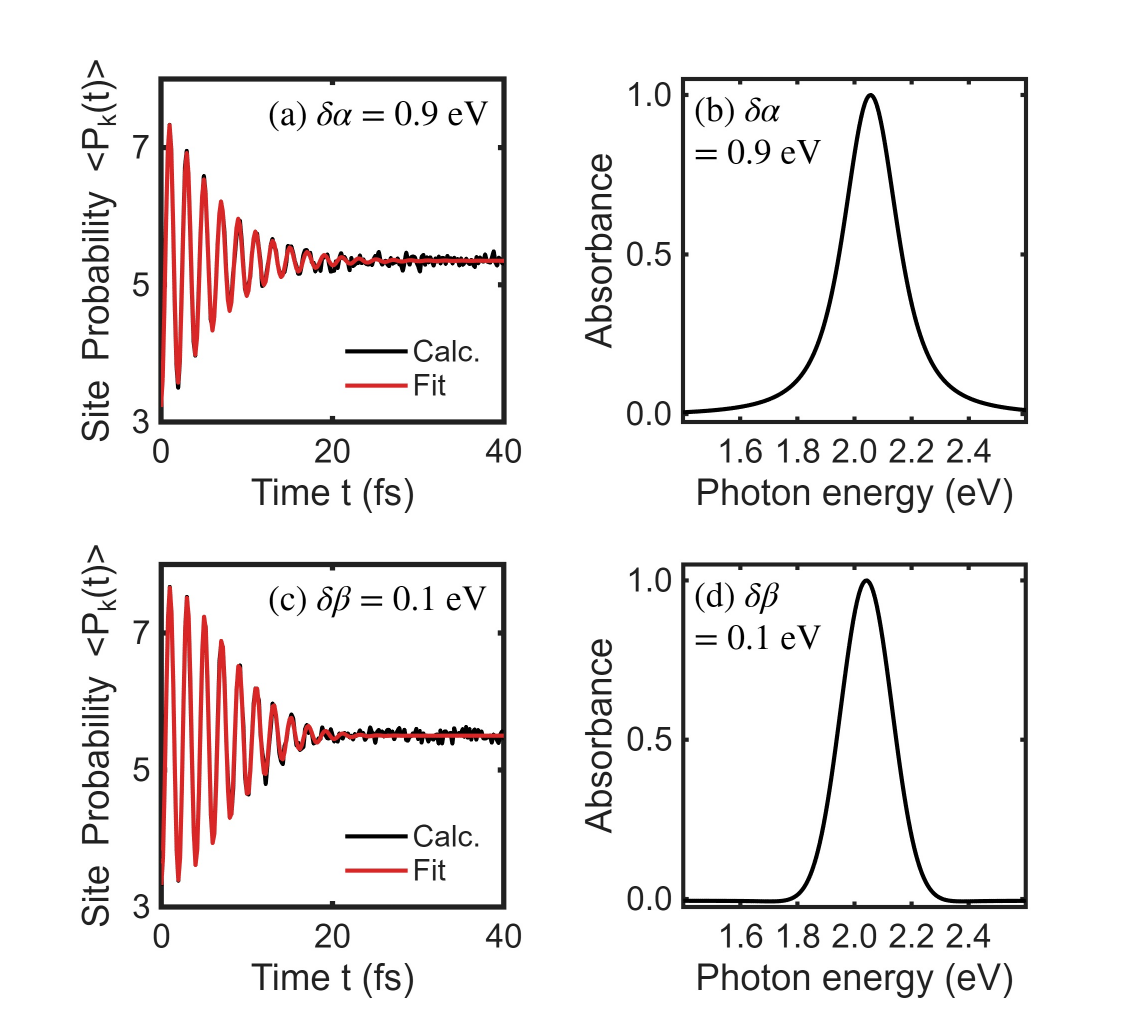}
\caption{ The electronic trajectory is fitted to Equation~\eqref{eq:stretched_damped_cosine} to obtain the corresponding absorption spectrum. Panels (a)--(b) show results for diagonal disorder with $\delta\alpha = 0.9$~eV, while the off-diagonal fluctuations are held fixed at $\delta\beta = 0.0$~eV. Panels (c)--(d) show the corresponding trajectory and absorption spectrum for off-diagonal disorder with $\delta\beta = 0.1$~eV and $\delta\alpha = 0.0$~eV. Compared to diagonal fluctuations, substantially smaller off-diagonal fluctuations lead to faster damping of the time-domain coherence and broader absorption features, highlighting the enhanced sensitivity of electronic coherence to torsional fluctuations along the polyene chain.}
\label{spectra}
\end{figure}

To compute absorption spectra, the ensemble-averaged site probabilities are obtained by averaging over \(J = 800\) independent trajectories:
\begin{equation}
\langle P_k(t) \rangle = \frac{1}{J}\sum_{j=1}^{J} P_k^{(j)}(t).
\end{equation}
In Figures~\ref{spectra}(a)--\ref{spectra}(b), the response of the polyene chain is first simulated with \(\delta\alpha = 0.9\)~eV and \(\delta\beta = 0.0\)~eV to demonstrate the sensitivity of the dephasing profile to diagonal fluctuations. The trajectory is fit to a stretched exponential to capture deviations from a Gaussian form:
\begin{equation}
f(t) = A \cos(\omega t)\,\exp\!\left[-(\sigma |t|)^{\xi}\right] + B,
\label{eq:stretched_damped_cosine}
\end{equation}
allowing the absorbance spectrum to be computed with minimal noise from the fitted dephasing profile via a Fourier transformation. The absorbance spectrum reveals significant Lorentzian character (\(\xi \approx 1.3\)) with a FWHM of 0.21~eV. In contrast, the calculations presented in Figures~\ref{spectra}(c)--\ref{spectra}(d) employ \(\delta\alpha = 0.0\)~eV and \(\delta\beta = 0.1\)~eV, resulting in a Gaussian spectrum (\(\xi \approx 2.0\)) with a FWHM of 0.20~eV. Although the standard deviation parameters for $\delta\alpha$ and $\delta\beta$ differ by a factor of 9, the absorption spectra displayed in Figure~\ref{spectra} exhibit similar line widths, as the dephasing process is most susceptible to off-diagonal fluctuations.

Fluctuations of the diagonal elements, $\alpha_k(t) = \bar{\alpha} + \delta\alpha_k(t)$, are understood to represent local shifts in site energies arising primarily from stochastic electrostatic (Coulombic) interactions between the polyene and its surrounding environment. For a delocalized molecular orbital $u$, the first-order change in its energy due to such fluctuations is given by
\begin{equation}
\delta E_u \approx \sum_{k=1}^{N} |C_{ku}|^2\,\delta\alpha_k,
\end{equation}
where $C_{ku}$ is the amplitude of orbital $u$ on site $k$. For extended orbitals in a linear polyene, the coefficients satisfy $|C_{ku}|^2 \sim 1/N$ over much of the chain, so that the energy shift becomes an average over many local perturbations. As a result, the magnitude of $\delta E_u$ scales as $\sim1/N$, reflecting the self-averaging effect of delocalization. Because the HOMO and LUMO in a linear polyene have similar spatial probability distributions, correlated shifts $\delta E_\mathrm{H}$ and $\delta E_\mathrm{L}$ lead to exchange narrowing:
\begin{equation}
\delta(E_\mathrm{L} - E_\mathrm{H}) \ll \delta E_\mathrm{L},\ \delta E_\mathrm{H}.
\end{equation}
Diagonal fluctuations predominantly generate such common-mode shifts of orbital energies rather than strong fluctuations of the energy gap, which results in comparatively weak dephasing.

In contrast, fluctuations of the off-diagonal elements, $\beta_{km}(t) = \bar{\beta} + \delta\beta_{km}(t)$, have a much stronger effect on decoherence because they directly perturb the electronic couplings that control delocalization along the chain. These couplings are particularly sensitive to the molecular geometry and are naturally modulated by structural motions such as torsional distortions of the conjugated backbone, which alter $\pi$-orbital overlap between neighboring carbon atoms. For a finite linear chain, the unperturbed H\"uckel eigenvalues corresponding to a tridiagonal matrix are
\begin{equation}
E_u = \alpha + 2\beta\cos\!\left(\frac{u\pi}{N+1}\right),
\qquad u = 1,2,\ldots,N,
\end{equation}
Therefore, a fluctuation $\delta\beta$ produces a direct first-order change in each orbital energy,
\begin{equation}
\delta E_u \approx 2\,\delta\beta\,\cos\!\left(\frac{u\pi}{N+1}\right).
\end{equation}
Unlike diagonal disorder, the fluctuations in the energy gaps are not suppressed by delocalization because they are generally uncorrelated among neighboring atoms.

Together, these considerations explain why the simulations exhibit a much more rapid loss of coherence when disorder is applied to the off-diagonal elements than when comparable disorder is applied to the diagonal elements. Diagonal fluctuations, which primarily reflect Coulombic solute--solvent interactions, tend to shift all delocalized orbitals together and therefore only weakly perturb the resonant energy gap that governs coherent motion. Off-diagonal fluctuations, which are naturally associated with structural distortions, directly disrupt electronic coupling and delocalization, producing strong phase randomization and rapid decoherence. From a pedagogical perspective, this distinction provides a clear physical message: the microscopic origin of environmental fluctuations plays a central role in determining the rate of decoherence and the manner in which optical spectral line shapes emerge in condensed-phase systems.

\section{Conclusion}

We have presented a computational and visualization-based framework for exploring electronic coherence and spectral line broadening in conjugated molecular systems using a time-dependent H\"uckel model. By combining explicit real-time propagation, analytic results for idealized polyene chains, and stochastic disorder to represent environmental interactions, the approach provides a physically transparent picture of how molecular-scale dynamics give rise to experimentally observable line shapes. The accompanying MATLAB implementation is designed for direct classroom use, enabling students to interactively explore coherent motion, decoherence, and absorption spectra across a wide range of model parameters. Example problem set are also provided in the Supplementary Material to facilitate direct implementation of these materials in instructional settings.

Beyond its practical utility, the framework emphasizes a central conceptual message: spectral line broadening arises from the progressive loss of electronic phase coherence, and the microscopic origin of environmental fluctuations plays a decisive role in determining how rapidly this coherence is lost. Electrostatic (diagonal) perturbations tend to shift molecular orbital energies in a correlated manner and weakly perturb coherent dynamics, whereas structural (off-diagonal) perturbations directly disrupt electronic coupling and delocalization, leading to efficient decoherence. By making these processes visible in real space and connecting them directly to frequency-domain spectra, the present approach provides students with an intuitive and rigorous bridge between quantum dynamics, molecular structure, and optical response.


\section*{Supplementary Material}
Supplementary Material is available as ancillary files in the arXiv version of this manuscript. It provides a complete analytic derivation of the coherent electronic dynamics on an \(N\)-site H\"uckel chain, including closed-form expressions for site probabilities and ensemble observables, along with additional mathematical details and instructional problem sets. The MATLAB script used to perform the numerical calculations is also provided.

\begin{acknowledgments}
This work is supported by the National Science Foundation under Grant Nos.~CHE-2247159.
\end{acknowledgments}

\section*{Conflicts of Interest }
The authors have no conflicts to disclose.

\section*{References}
\bibliography{polyenebibliography}

\end{document}